\documentclass[showpacs,preprintnumbers,amsmath,amssymb,prl,twocolumn]{revtex4}
\usepackage{graphicx}
\begin{document}

\title{Unusual conductance collapse in one-dimensional
quantum structures}

\author{K~J~Thomas}
\email{kjt1003@cam.ac.uk}
\author{D~L~Sawkey}
\email{sawkey@mailaps.org}
\author{M~Pepper}
\author{W~R~Tribe}
\altaffiliation{Present address:  TeraView Ltd., Cambridge, CB4 0WG, United Kingdom}
\author{I~Farrer}
\author{M~Y~Simmons}
\altaffiliation{Present address: School of Physics, The University of New South Wales, Sydney 2052, Australia}
\author{D~A~Ritchie}
\affiliation{Cavendish Laboratory, Madingley Road, Cambridge, CB3
0HE, United Kingdom}

\date{March 19, 2004: Published: J. Phys.: Condens. Matter \textbf{16}, L279--L286 (2004)}

\begin{abstract}

We report an unusual insulating state in one-dimensional quantum
wires with a non-uniform confinement potential.  The wires consist
of a series of closely spaced split gates in high mobility
GaAs/AlGaAs heterostructures. At certain combinations of wire
widths, the conductance abruptly drops over three orders of
magnitude, to zero on a linear scale.  Two types of collapse are
observed, one occurring in multi-subband wires in zero magnetic
field and one in single subband wires in an in-plane field.  The
conductance of the wire in the collapse region is thermally
activated with an energy of the order of 1~K. At low temperatures,
the conductance shows a steep rise beyond a threshold DC
source-drain voltage of order 1~mV, indicative of a gap in the
density of states. Magnetic depopulation measurements show a
decrease in the carrier density with lowering temperature. We
discuss these results in the context of many-body effects such as
charge density waves and Wigner crystallization in quantum wires.
\end{abstract}


\maketitle

Electron-electron interactions profoundly modify transport in one
dimension. For short-range interactions, a 1D system is described
as a Luttinger liquid (LL) with gapless excitations
\cite{tomanaga50,luttinger63, haldane81}. In the presence of
scattering, the conductance of an LL with repulsive interactions
is predicted to vanish in a power-law fashion as temperature $T
\rightarrow 0$ \cite{kane92c}.  For long-range interactions at
sufficiently low densities a single subband wire is predicted to
form a 1D Wigner crystal \cite{glazman92,schulz93}. Theoretical
studies also point to the existence of a charge density wave order
in quantum wires \cite{kane92c,glazman92,larkin}.  In the case of
multi-subband wires, an inter-subband charge density wave can be
stabilized, which can give rise to an insulating state due to
inter-subband backscattering, or superconducting instabilities for
Cooper scattering \cite{aoki94,starykh}. Unlike a Luttinger
liquid, such states are expected to be gapped in quantum wires at
low temperatures \cite{starykh}.

Observation of interaction effects requires that the length $L$ of
the wire be much larger than the thermal coherence length $L_T =
\hbar v_F / k_B T$, where $v_F$ is the Fermi velocity.  In shorter
wires, interactions are renormalized by the non-interacting leads
\cite{maslov95a,safi95,pmarenko95} and transport is broadly
understood within the picture of a non-interacting electron gas,
although interaction effects such as the 0.7 structure
\cite{thomas96a}, enhanced $g$-factors \cite{thomas96a,daneswar}
and the high-field 0.7 analogue structures \cite{graham03} are
still present. Fabricating clean, long 1D wires is a technological
challenge \cite{tarucha95,liang00}. However, in the presence of
interactions, disorder and backscattering play an important role
\cite{kane92c,apel82,ogata94}. Power-law behaviour in temperature
and voltage, expected for a dirty Luttinger liquid, has been
measured in carbon nanotubes \cite{bockrath99} and in GaAs wires
\cite{tarucha95}.

We have fabricated a long quantum wire system formed by the series
addition of two or three closely spaced shorter wires defined in
high mobility two-dimensional electron gases (2DEG), allowing the
controlled modulation of the confining potential.  By varying the
widths of the wires independently, we observe a collapse in the
conductance of greater than three orders of magnitude. DC bias
measurements show that a gap is present in the density of states.
Perpendicular magnetic field measurements show a reduction in
carrier concentration as the temperature is lowered. In a parallel
field the collapse is enhanced.  Although there are differences,
such behaviour resembles theoretical predictions of a 1D Wigner
crystal \cite{glazman92,schulz93} and charge density wave order
\cite{kane92c,glazman92,larkin}.

We present results observed in three samples. Samples A and B
consist of three split gates in series, defined in a single
GaAs/AlGaAs quantum well. The 2DEG formed 300 nm below the surface
has a mobility $\mu=3.4 \times 10^6$~cm$^2$V$^{-1}$s$^{-1}$ and
density $n=2.4 \times 10^{11}$~cm$^{-2}$ at 1.5~K\@.  Figure~1(a)
inset shows a schematic of samples A and B, which have a short
wire of lithographic length $L=0.4$~$\mu$m between two long wires
of $L=3$~$\mu$m. All wires have a width $W=1.2$~$ \mu$m and the
separation between adjacent wires is $S=0.4$~$ \mu$m. When the
gates are negatively biased, the projected dimensions of the wires
in the 2DEG change due to electrostatic depletion; $L$ increases
and $W$ and $S$ decrease, resulting in an inhomogeneous quantum
wire system longer than 7.2 $\mu$m. The results reported here do
not qualitatively differ even when the central short wire is left
grounded. The inhomogeneity of the potential at the junction
between the long wires acts as a scattering centre
or pinning source.  The role of the central short wire is to modulate
the potential at this region.

Sample C is defined in a single heterojunction with the 2DEG
100~nm below the surface, with $\mu=3.6 \times
10^6$~cm$^2$V$^{-1}$s$^{-1}$ and $n=3.3 \times 10^{11}$~cm$^{-2}$
after brief illumination with a red LED. The sample (figure 4
inset) consists of a long wire of $L=2$~$\mu$m and $W=0.8$~$\mu$m
between two short wires of $L=0.3$~$\mu$m and $W=0.4$~$\mu$m.  The
separation between the wires is 0.25~$\mu$m. Only the long wire
and one of the short wires need to be defined to see the effects
reported here. For all the samples, the inter-wire spacing is less
than the width of the wires, implying that there is no 2DEG
between the wires when defined.  Standard two terminal conductance
measurements are made with the sample attached to the mixing
chamber of a dilution refrigerator.  The excitation voltage is
$\leq 10$~$\mu$V at a frequency of 31 Hz. The data presented here
are corrected for a series resistance of 1~k$\Omega$, unless
otherwise mentioned.

Two types of conductance collapse are observed, both occurring as the
width of one of the wires is swept while the others are held
constant.  Type 1 conductance collapses (CC1) occur in zero magnetic
field and show a hysteresis in gate voltage, and type 2 (CC2) occurs
only in the presence of a parallel magnetic field, but is reproducible
in both directions of gate sweep.

Figure 1(a)
\begin{figure}
\begin{center}
\includegraphics[width=8cm]{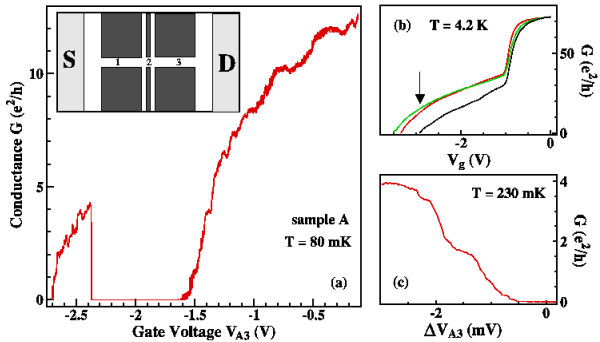}
\end{center}
\caption{(a) Conductance $G(V_{g})$ characteristics of sample A
showing collapse CC1 in the upward sweep of wire A3 whilst wires
A1 and A2 are held at -$2.91$ and 0~V respectively. {\it Inset}:
Schematic of samples A and B, showing three split gates in series.
Dimensions are given in the text. (b) $G(V_g)$ data (not corrected
for series resistance) of the long wires showing definition and
pinch-off characteristics: Green trace for A1, red trace for A3,
and black trace for sweeping A1 and A3 simultaneously. (c) Detail
of the collapse (abrupt drop in $G$) in a slow sweep showing a
plateau at $2e^2/h$ (in the same cool down, but a different
measurement from (a)). The $x$-axis origin is offset by
$2.028$~V.}
\end{figure}
shows a typical example of CC1 at 0.1~K\@, where wire 1 of sample
A (A1) was fixed at $-2.91$~V corresponding to several occupied 1D
subbands, and $G$ is measured as a function of $V_{A3}$.
Figure~1(b) shows the $G(V_g)$ characteristics of the two wires A1
and A3, measured independently as well as in series.
The arrow indicates the voltage (-2.91~V) at which wire A1 is
fixed for the measurement in figure~1(a).  When A1 is biased at
this voltage, the series combination of A1 and A3 together
pinch-off at $V_{A3}=-2.7$~V, as shown in figure 1(a). As $V_{A3}$
was swept towards 0~V, the two-terminal conductance $G \equiv
dI/dV$ rose to a value close to $4e^2/h$ and abruptly collapsed to
zero (within the measurement noise level) at $V_{A3}=-2.37$~V. The
drop in the conductance is more than three orders of magnitude. On
continuing the voltage sweep, the conductance increases gradually
from zero for $V_{A3}>-1.65$~V\@. In a slow sweep of $V_{A3}$
(different measurement), shown as figure 1(c), the collapse edge
shows a plateau at $2e^2/h$. At low temperatures, the zero
conductance state is stable and lasts for several days until the
experiment is terminated.  At higher temperatures, the conductance
drops to a value above zero, but the drop is still abrupt. No such
collapse has been observed in a single wire.

To highlight the reproducibility of CC1 in the up sweeps and the
absence of it in the down sweeps, we show in figure~2(a)
\begin{figure}
\begin{center}
\includegraphics[width=8cm]{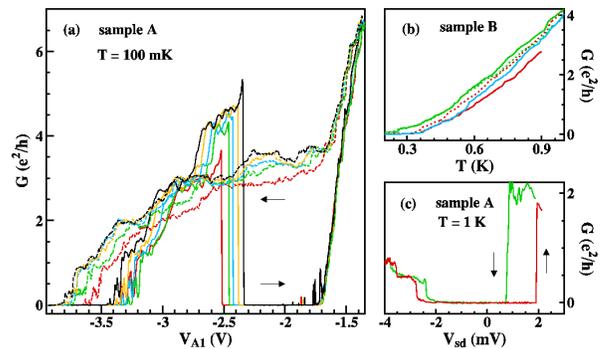}
\end{center}
\caption{Five sets of up (solid line) and down (dashed line)
sweeps showing reproducibility of collapse CC1 in the up sweeps of
$G(V_g)$. There is no collapse in the down sweeps. Wire A1 is
swept whilst A3 is held constant (-2.8 V) in contrast to figure
1(a). The arrows indicate direction of sweep. For a more negative
pinch-off voltage, the collapse is triggered at a less negative
gate voltage and from a higher $G$\@. (b) $G(T)$ of sample B in
the collapse region (with fixed gate voltages), uncorrected for
series resistance, showing reproducibility in temperature sweeps.
Each colour represents a temperature cycle, solid and dotted lines
are cooling and heating, respectively. (c) DC bias measurements
showing hysteresis and a gap in subsequent sweeps in sample A with
$V_{A1}=-1.87$~V, $V_{A2}= 0$~V, and $V_{A3}=-3$~V.}
\end{figure}
five sets of measurements taken under the same conditions.  Here
$V_{A3}$ is fixed at $-2.8$~V and $V_{A1}$ is swept.  The
conductance collapse is only observed in the up sweeps (solid
lines) for $-2.5$~V $<V_{A1}<-1.7$~V, whereas the conductance for
the down sweeps (dashed lines) remains finite, close to $3e^2/h$,
for the same range of $V_{A1}$. The large hysteresis in the
characteristics of the up and down voltage sweeps is a feature of
CC1. We note that for either direction of sweep, the
characteristics to the left of the collapse ($V_{A1}<-2.2$~V) in
figure~2(a) are not exactly reproduced in all the traces, but to
the right of the collapse ($V_{A1}>-1.7$~V) they are nearly identical.
A change in pinch-off voltage accompanies a change in collapse
voltage for different sweeps. Traces with a more negative
pinch-off voltage have a conductance collapse at a less negative
voltage and from a higher conductance value.  With gate voltages
fixed in the collapse region for sample B, figure~2(b) shows the
reproducibility of a typical collapse state in several temperature
cycles between 0.2 and 1~K\@.

Figure 2(c) shows the DC source-drain bias $G(V_{sd})$
measurements taken in the collapse region for sample A at 1~K\@.
At $V_{sd}=0$, the collapse state was stable at $G=0$ for
$T<1.4$~K. As $|V_{sd}|$ increases, we observe a sudden jump in
the conductance from zero to a nearly constant value, indicative
of a gap in the density of states or depinning. The gap shows
hysteresis and a jump to a different $G$ value for $+$ and
$-V_{sd}$. The gap, 3--5~mV, is much greater than both the highest
temperature at which any collapse is observed and the activation
temperature of 5.5 K measured in this cooldown.

Transverse magnetic field measurements show that the collapse is
accompanied by a reduction in the 1D carrier density with
decreasing temperature. The effect of a transverse magnetic field
$B_z$ on 1D wires is to increase the energy spacing of the 1D
subbands and cause them to pass progressively through the Fermi
energy \cite{berggren86}. The energies of the magnetoelectric 1D
levels are determined by the confinement provided by the gate
voltages and the magnetic field. At a constant gate voltage, as in
our case, magnetic depopulation causes $G$ to decrease as
$|{B_z}|$ is increased, with plateaus of quantized conductance
appearing as the energy levels are forced through the Fermi
energy. When the cyclotron diameter is greater than the wire width
the levels are hybrid magnetoelectric levels, and when the
cyclotron diameter is much less than the width the 1D subbands are
equivalent to 2D Landau levels \cite{berggren86}.  From the
filling factor $\nu$ of Landau levels, the electron density
$N_{2D}=\nu (eB_z/h)$ can be extracted. Figure 3(a)
\begin{figure}
\begin{center}
\includegraphics[width=8cm]{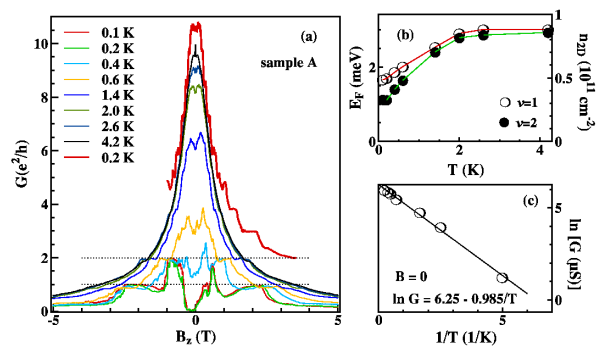}
\end{center}
\caption{(a) Magnetoconductance $G(B_z)$ at various temperatures
in the collapse region, at fixed gate voltages $V_{A1}=-1.55$~V,
$V_{A2}=-1.66$~V, and $V_{A3}=-2.7$~V taken in a different
cooldown from figures 1 and 2. The thick red trace, offset
vertically by $e^2/h$, is taken just outside the collapse (by
sweeping A2 to 0~V) at 0.2~K, and shows agreement with the trace
at 4.2~K in the collapse region. This indicates that the carrier
density is restored on heating in the collapse regime. There is a
slight asymmetry with $+/-B_z$ in the characteristics at low
temperatures, and at 4.2~K a sharp peak occurs reproducibly at
zero field. The data are uncorrected for series resistance. (b)
Carrier density measured from the $B_z$ positions of the filling
factors $\nu=1$ (open circles) and $\nu=2$ (closed circles) from
(a). (c) Arrhenius plot of $G$ at $B_z=0$ from (a).}
\end{figure}
shows the magnetoconductance $G(B_z)$ at various temperatures for
sample A in the collapse region with fixed gate voltages on each
of the three wires.  As $T$ decreases, $G$ at $B_z=0$ decreases
and there is a decrease in the magnetic field required to
depopulate a particular subband. For example, at 4.2~K the
$2e^2/h$ conductance plateau occurs at $B_z = 2$~T, but at 0.4~K
the plateau shifts to 1~T, corresponding to a 50\% reduction in
the (2D) carrier concentration, or an equivalent drop in the Fermi
energy, shown in figure~3(b).

Starting from the collapse region, a perpendicular field of 200~mT
is sufficient to restore conductance. For $T<0.2$~K, the
conductance first rises to a value close to $2e^2/h$ with
increasing $|B_z|$.  The thick red trace in figure 3(a) shows
$G(B_z)$ measured at 0.2~K but outside the collapse region. The
depopulation trace is almost identical with that measured in the
collapse region at 4.2~K (black trace), indicating that the
carrier density is restored on heating and showing that the drop
in carrier concentration is a property solely of the long wire
structure formed from the two in series. Figure 3(c) shows an
Arrhenius plot of $G(B_z =0)$ from figure 3(a), with an activation
temperature of 1~K.

The second type of collapse, CC2, observed in the presence of an
in-plane magnetic field, is reproducible in both directions of
gate voltage sweep. Figure 4
\begin{figure}
\begin{center}
\includegraphics[width=8cm]{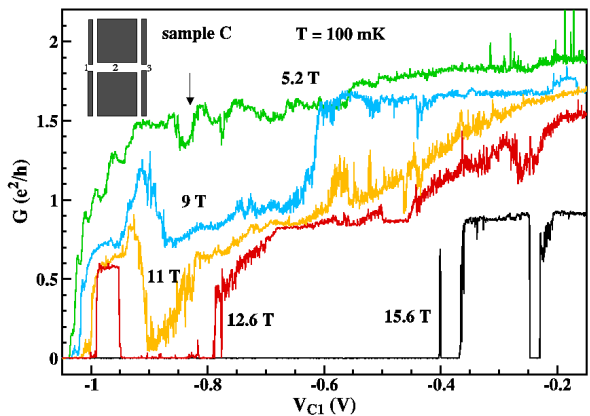}
\end{center}
\caption{$G$ at $T=100$~mK of sample C as a function of $V_{C1}$
at different in-plane magnetic fields, and a fixed voltage on wire
C2 such that $G\approx 2e^2/h$ at $B=0, V_{C1}=0$. The conductance
collapse is triggered at the region marked by the arrow. At
$B_{\parallel}=5.2$~T (green trace) there is no spin-split
plateau. As $B_{\parallel}$ increases the valley at the arrow
descends to $G=0$. The valley expands to a wider range of $V_{C1}$
at 15.6~T, and new regions of collapse appear, for example at
$V_{C1}\sim-0.23$~V.  A clear spin-split plateau close to $e^2/h$
can be observed at 12.6~T at $-0.7$~V$<V_{C1}<-0.45$~V\@.  {\it
Inset}: Schematic of sample C (see text for dimensions).}
\end{figure}
shows five $G(V_{C1})$ traces in sample C at various in-plane
magnetic fields, $B_{\parallel}$, applied parallel to the wires.
The misalignment of the field was less than 1$^{\circ}$ out of the
2DEG plane. For this measurement $V_{C2}$ is such that
$G_{C2}=2e^2/h$ at $B_{\parallel}=0$. Strong fluctuations are
present in the traces. As $B_{\parallel}$ increases, a small
valley at $B_{\parallel}=5.2$~T marked by an arrow drops to zero
conductance at $B_{\parallel}=12.6$~T, and the width of the
collapse region increases.  At $B_{\parallel} = 15.6$~T, the
valley expands and a new region of collapse is seen at $V_{C1}
\sim -0.23$~V\@. The conductance drop is typically $e^2/h$ and the
gate voltage width of the collapse was within $k_BT$ broadening of
zero.  The collapse was continuously reproduced in both sweep
directions over 100 times with no appreciable change in the
traces. Shifting the short wire C1 laterally has no appreciable
effect.  Collapse type CC1 was also observed in this sample, but
reproducibility was low.  It is unknown whether the reproducibility of CC2 in
the up anddown sweeps, in contrast to CC1, is associated with the lifting of
the spin degeneracy of the electrons.

Traces of CC2 at different temperatures at 12.6~T are shown in
figure~5.
\begin{figure}
\begin{center}
\includegraphics[width=8cm]{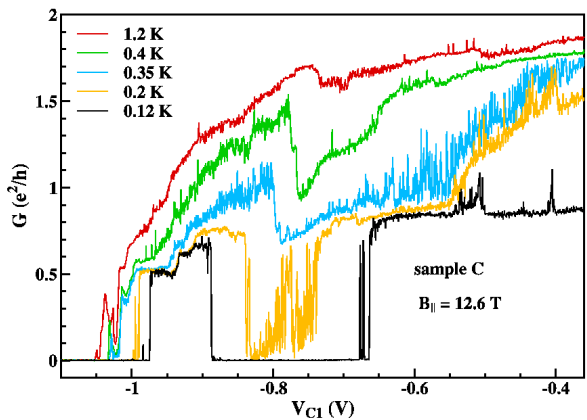}
\end{center}
\caption{Temperature dependence at 12.6~T of the conductance
collapse CC2.  There is a small drift in the characteristics at
0.12~K from figure~4. The collapse is absent at 0.2~K, and at
0.35~K the spin-split plateau disappears and $G$ rises showing
strong fluctuations. At 1.2~K, $G$ rises by $e^2/h$ outside the
collapse region and $1.7(e^2/h)$ in the collapse region.}
\end{figure}
At 0.12~K the collapse is well defined, and outside the collapse
region the conductance is suppressed to $e^2/h$.  At 1.2~K strong
fluctuations set in, the collapse is removed, and there is no
spin-split plateau at $e^2/h$.  The disappearance of the
spin-split plateau and the associated rise in $G$ at high
temperatures resemble our previous observations on the 0.7
structure \cite{thomas00}.

Various mechanisms can be responsible for small drops in
conductance. Resonances, resulting from defects, impurities, or
modifications of the exit regions; universal conductance
fluctuations \cite{BeenRev}; and for double constrictions,
anti-resonances due to mode matching \cite{blaikie} are possible.
These mechanisms, however, cannot explain collapses of the
magnitude observed here, nor the hysteresis and temperature
dependences. Localization of carriers by disorder is inconsistent
with figure 2(c).  No anomalies were reported in previous studies
of short wires in series \cite{wharam88b,kouwen89}.

Theories exist which predict the existence of an insulating state
in 1D wires.  In the presence of any scattering the conductance of
a Luttinger liquid is expected to vanish as a power law in
temperature and voltage. The observed $G-V_{sd}$ behaviour (figure
2(c)) is inconsistent with such a mechanism. An insulating state
may result from the pinning of a Wigner crystal or a charge
density wave by a barrier potential. The $G-V_{sd}$ relation and
thermally activated behaviour support such mechanisms
\cite{glazman92}. Our 2D density immediately before the collapse
is $3\times10^{10}$~cm$^{-2}$. For a width of 0.3~$\mu$m, equal to
the cyclotron diameter at 0.2 T, where the collapse is removed in
figure 3(a), we estimate $n_{1D} \approx 1\times10^6$~cm$^{-1}$ or
$n_{1D} a_B \sim 1$, $a_B$ being the Bohr radius.  At such low
densities a 1D Wigner crystal \cite{glazman92,schulz93} in single
subband wires is expected, and may show a pinning energy greater
than the thermal activation energy, as we observe.  A trapped
superconducting state \cite{lambert98} also cannot be excluded.
For multiple subbands, an inter-subband charge density wave
\cite{starykh} could open a gap at the Fermi
level due to inter-subband backscattering, leading to an
insulating state. However, the drop in Fermi energy with
temperature further suggests the formation of a state below the 1D
subbands as $T \rightarrow 0$. We also note the suppression of the
collapse with a small perpendicular field, which suggests that the
interaction between $+$k and $-$k electrons is important.

In summary we have observed an unusual collapse in the conductance
of serially connected 1D quantum wires.  Two types of collapse are
observed, both of which have the defining feature of zero
conductance in a gate voltage region where a finite conductance is
expected.  The mechanism is unknown but may result from an
inter-subband charge density wave or a 1D Wigner crystallization.
Further experimental and theoretical work is necessary to
understand the phenomenon.

We acknowledge EPSRC for funding this work, and KJT
thanks the Royal Society for the Eliz Challenor Research Fellowship.
We thank Alan Beckett and Pete Flaxman for technical assistance.

\bibliographystyle{plain}
\end{document}